\documentclass[prb,twocolumn,showpacs]{revtex4-1}

\usepackage{graphicx,bm}

\newcommand{\bfr}{{\bf r}}

\newcommand{\qp}{q_{||}}
\newcommand{\bfqp}{{\bf q}_{||}}
\newcommand{\rp}{r_{||}}
\newcommand{\bfrp}{{\bf r}_{||}}
\newcommand{\kp}{k_{||}}
\newcommand{\bfkp}{{\bf k}_{||}}
\newcommand{\rhop}{\rho_{||}}
\newcommand{\bfrhop}{\bm{\rho}_{||}}
\newcommand{\ua}{\uparrow}
\newcommand{\da}{\downarrow}

\newcommand{\PGG}{^{\rm PGG}}
\newcommand{\ISTLS}{^{\rm ISTLS}}

\begin{document}

\title{Three- to two-dimensional crossover in time-dependent density-functional theory}

\author{Shahrzad Karimi}
\affiliation{Department of Physics and Astronomy, University of Missouri-Columbia,
Columbia, Missouri, 65211}

\author{Carsten A. Ullrich}

\affiliation{Department of Physics and Astronomy, University of Missouri-Columbia,
Columbia, Missouri, 65211}

\date{\today }

\begin{abstract}
Quasi-two-dimensional (2D) systems, such as an electron gas confined in a quantum well,
are important model systems for many-body theories. Earlier studies of the
crossover from 3D to 2D in ground-state density-functional theory showed that local
and semilocal exchange-correlation functionals which are based on the 3D electron gas
are appropriate for wide quantum wells, but eventually break down as the
2D limit is approached. We now consider the dynamical case and study the performance
of various linear-response exchange kernels in time-dependent density-functional theory.
We compare approximate local, semilocal and orbital-dependent exchange kernels,
and analyze their performance for inter- and intrasubband plasmons as the quantum wells approach
the 2D limit. 3D (semi)local exchange functionals are found to fail
for quantum well widths comparable to the 2D Wigner-Seitz radius $r_s^{\rm 2D}$, which
implies in practice that 3D local exchange remains valid in the quasi-2D dynamical regime for typical quantum well parameters,
except for very low densities.
\end{abstract}


\maketitle

\section{Introduction}

The key concept of density-functional theory (DFT)\cite{Hohenberg1964} is that all electronic many-body systems can be
uniquely characterized by their electron density $n(\bfr)$. The density can be obtained in principle exactly
via the Kohn-Sham equation (here and in the following we use atomic units),\cite{Kohn1965}
\begin{equation}\label{1}
\left[ -\frac{ \nabla^2}{2} + v_0(\bfr) + v_{\rm H}[n](\bfr) + v_{\rm xc}[n](\bfr)\right]\varphi_j(\bfr)
= \varepsilon_j \varphi_j(\bfr) ,
\end{equation}
where $v_0(\bfr)$ is a given external potential,
$v_{\rm H}[n](\bfr) = \int d^3r' \: n(\bfr')/|\bfr - \bfr'|$
is the Hartree potential, and $v_{\rm xc}[n](\bfr)$ is the exchange-correlation (xc) potential. The density
is obtained from the self-consistent solution of Eq. (\ref{1}) as $n(\bfr) = \sum_{j=1}^N |\varphi_j(\bfr)|^2$,
where $N$ is the number of electrons, and all physical observables follow therefrom.

The xc potential is defined as the functional derivative  $v_{\rm xc}[n](\bfr) = \delta E_{\rm xc}[n]/\delta n(\bfr)$.
The xc energy $E_{\rm xc}[n]$ is a {\em universal} functional of the density: this means that there is one
and only one exact density functional of the xc energy that is valid for {\em all} electronic systems with
a given form of the electron-electron interaction, for any $N$. If this exact xc functional were known, it would give
exact ground-state results, via Eq. (\ref{1}), for all conceivable forms of matter, including atoms, molecules,
and periodic or non-periodic solids.

In real matter, $v_0(\bfr)$ consists of the Coulomb potentials of
positively charged atomic nuclei.
But the universality of $E_{\rm xc}[n]$ and $v_{\rm xc}[n](\bfr)$ extends beyond real matter, and
includes all mathematically reasonable forms of $v_0(\bfr)$, whether they exist in nature
or not. In particular, it includes systems of lower dimensionality, for instance electrons confined in
a two-dimensional (2D) plane.\cite{footnote}

A stringent test for approximate xc functionals is their performance during a
dimensional crossover. The cross\-over from 3D to 2D has been previously studied in the  DFT literature. \cite{Pollack2000,Kim2000,Constantin2008a,Constantin2008b} It was found that local and semilocal
functionals such as the local-density approximation (LDA) and generalized gradient approximations (GGAs)
fail badly at this task. To see this, consider the LDA exchange energy
\begin{equation}\label{2}
E_{\rm x,3D}^{\rm LDA}[n] = -\frac{3}{4}\left(\frac{3}{\pi}\right)^{1/3} \int d^3r\:  n(\bfr)^{4/3} \:.
\end{equation}
What happens if we try to evaluate $E_{\rm x,3D}^{\rm LDA}[n]$ for a 2D system? Let the density be
$n_{\rm 2D}(\bfr) = n(\bfrp) \delta(z)$,
where $\bfrp = (x,y)$ denotes a 2D position vector. Using the delta function in the form
$\delta(z) = \lim_{\epsilon \to 0^+} (4\pi\epsilon)^{-1/2}e^{-z^2/4\epsilon}$,
one finds
\begin{equation}\label{3}
E_{\rm x,3D}^{\rm LDA}[n_{\rm 2D}] = \lim_{\epsilon \to 0^+} \frac{3^{11/6}}{4^{5/3}\sqrt{\pi} \epsilon^{1/6}}
\int d^2 \rp \:  n(\bfrp)^{4/3} \:.
\end{equation}
This clearly shows that the 3D form of the LDA exchange energy diverges in the 2D limit, instead of approaching
the proper form of the 2D LDA,\cite{GV}
\begin{equation}\label{4}
E_{\rm x,2D}^{\rm LDA}[n] = - \frac{4}{3}\sqrt{\frac{2}{\pi}} \int d^2 \rp\:  n(\bfrp)^{3/2} \:.
\end{equation}
All standard 3D GGAs will exhibit a similar divergence in the 2D limit.

To capture the 3D-2D crossover correctly, nonlocal xc functionals are needed. Some improvement
over LDA and GGAs can be achieved with meta-GGA and hyper-GGA xc functionals, \cite{Constantin2008a,Constantin2008b}
but only fully nonlocal xc functionals such as the average density approximation \cite{Kim2000}
or the inhomogeneous STLS \cite{Dobson2002,Dobson2009} show a proper behavior as the 2D limit is approached.

In this paper, we extend the study of the dimensional crossover into the domain
of time-dependent density-functional theory (TDDFT).\cite{Runge1984,Ullrich2012,Ullrich2014}
However, we will not explore the full dynamical range of TDDFT, which allows one to
study electronic systems under the influence of arbitrary external time-dependent potentials, $v(\bfr,t)$;
instead, we will limit ourselves to the linear-response regime and consider
electronic excitation energies.\cite{Casida1995,Petersilka1996} Furthermore, in this paper
we will only consider exchange, but not correlation effects.

The main questions are the following.
What characteristic effects or signatures occur in the excitation spectrum of a system as it crosses over from three to
two dimensions, and how will the expected failure of LDA and GGA manifest itself? Will the breakdown
be as drastic as in ground-state DFT, or will it perhaps be less severe, under some circumstances?
How do nonlocal orbital functionals perform under the 3D-2D crossover?

Apart from the inherent fundamental interest, there are important practical reasons
that motivate such a study. Quasi-2D\cite{2Dfootnote} electron gases (2DEGs)
can be prepared in very high quality along interfaces and in heterostructures of a wide range of materials
(most notably semiconductors and oxides), with many practical applications.\cite{Davies,Harrison}
It is important to be able to model the electronic structure and dynamics in these systems accurately
and numerically efficiently. Since no DFT method beats the LDA in terms of simplicity and efficiency, one would like
to know whether the 3D LDA is reliable in the quasi-2D regime, and under what circumstances it
starts to fail. We will answer these questions in the following.

\begin{figure}
\begin{center}
\includegraphics[angle=0, width=8.5cm]{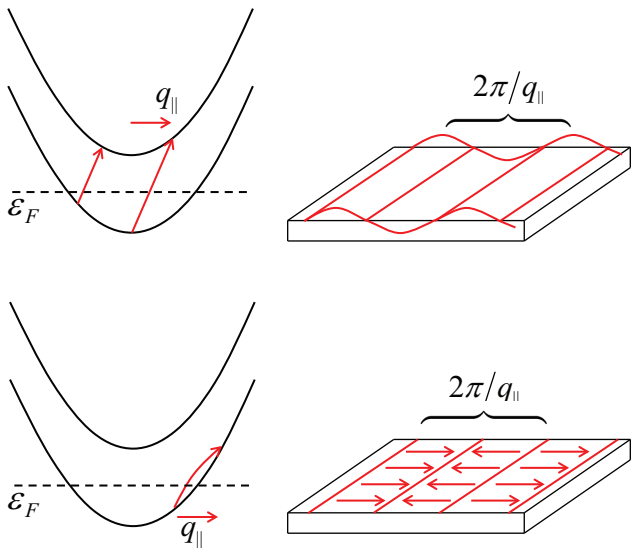}
\end{center}
\caption{Illustrations of intersubband (top) and intrasubband (bottom) plasmon excitations with wavevector $\qp$
in a quantum well with conduction band Fermi level $\varepsilon_F$ in the lowest subband.
Intersubband plasmons involve collective transitions between two subbands, leading to density oscillations of
the quasi-2D electron system perpendicular to the
quantum well plane. Intrasubband plasmons (collective transitions within the lowest subband)
are characterized by density oscillations and currents flowing along the plane.}
\label{fig1}
\end{figure}

Figure \ref{fig1} illustrates the two types of collective excitations that we will study in this paper.
In a quantum well, electrons are free to move in the plane, but the levels are quantized into subbands
due to quantum confinement perpendicular to the plane. Intersubband plasmons involve transitions
from occupied to empty subbands; since different subbands have different envelope functions, this implies
density oscillations {\em perpendicular} to the well plane. By contrast, intrasubband plasmons involve
transitions within a subband; the accompanying currents and density oscillations are {\em parallel} to the plane.
We will study what happens to these excitations as the quantum well becomes more and more narrow, approaching
the strictly 2D limit.

This paper is organized as follows. In Section II we discuss the necessary theoretical background: we
introduce our quantum well model, review the TDDFT linear-response formalism  for collective excitations in quantum wells,
and list various exchange functionals. In Section III we present our results, and
Section IV gives conclusions. Some technical details are given in the Appendix.

\section{Theoretical background}

\subsection{Quantum well model}

We consider n-doped semiconductor quantum wells of width $L$ in which the electrons are confined along the $z$ direction and free
to move in the $x-y$ plane.  The number
of electrons per unit area (the sheet density) is denoted by $N_s$. In the following, we assume that
the material of the quantum well is GaAs, with effective mass $m^* = 0.067 m$
and effective charge $e^*=e/\sqrt{13}$ ($m$ and $e$ are the free electron mass and charge). We choose
units in which $e^* = m^* = \hbar=1$. The effective Hartree unit of energy is
10.8 meV; the effective Bohr radius is 103 \AA.

The quantum well is assumed to be confined within infinitely high barriers at $z=0$ and $z=L$.
We further assume that the solutions of the Kohn-Sham equation for the quantum well envelope functions\cite{Davies,Harrison}
have the standard particle-in-a-box form,
\begin{equation} \label{box}
\varphi_j(z) = \sqrt{ \frac{2}{L}} \sin\left(\frac{j\pi z}{L}\right) ,\quad j=1,2,3,\ldots \:,
\end{equation}
with Kohn-Sham energies
\begin{equation} \label{II.28}
\varepsilon_j = \frac{1}{2} \left(\frac{j\pi}{L}\right)^2 .
\end{equation}
The Kohn-Sham potential $v_s(z) = v_{\rm ext}(z) + v_{\rm H}(z) + v_{\rm xc}(z)$ that gives rise to these
solutions is an infinitely deep square-well potential. This means that for each $L$ and $N_s$
the external quantum well potential $v_{\rm ext}(z)$ is chosen such that, if added to
the Hartree and xc potentials $v_{\rm H}(z)$ and $v_{\rm xc}(z)$, the resulting sum is a constant for $0<z<L$.
Thanks to the Hohenberg-Kohn theorem,\cite{Hohenberg1964} a unique choice of such a $v_{\rm ext}(z)$ is always possible in
principle; further details of the ground-state potentials do not need to be specified in the following.

We emphasize that the particle-in-a-box form of the Kohn-Sham eigenstates is only a matter
of convenience, and does not lead to a loss of generality of the results of the 3D-2D crossover that we study
in this paper.

The ground-state density in the well is given by
\begin{equation}
n_0(z) = \frac{1}{\pi} \sum_{j \atop \varepsilon_j < \varepsilon_F} \varphi_j^2(z)(\varepsilon_F - \varepsilon_j) \:.
\end{equation}
To determine the Fermi energy $\varepsilon_F$, we integrate the density over $z$:
\begin{equation}
\int_0^L dz n_0(z) = N_s = \frac{1}{\pi} \sum_{j=1}^{N_{\rm occ}}(\varepsilon_F - \varepsilon_j) \:,
\end{equation}
where $N_{\rm occ}$ is the number of occupied subbands. Hence,
\begin{equation}\label{II.31}
\varepsilon_F = \frac{\pi N_s}{N_{\rm  occ}} +  \frac{1}{N_{\rm occ}}\sum_{j=1}^{\rm N_{\rm occ}} \varepsilon_j,
\end{equation}
and $N_{\rm occ}$ is fixed by requiring $\varepsilon_{N_{\rm occ}} < \varepsilon_F < \varepsilon_{N_{\rm occ}+1}$.

\subsection{Excitations within linear-response TDDFT}

In the following, we are interested in the frequency-dependent spin-density response in a quantum well.
Because of the translational symmetry in the $x-y$ plane, we Fourier transform with respect to the
in-plane position vector $\bfrp = (x,y)$; this introduces the in-plane wavevector $\bfqp$. The TDDFT
linear-response equation \cite{Gross1985} then becomes
\begin{equation} \label{L.3.1}
n_{1\sigma}(\bfqp,z,\omega) = \int dz' \chi_{s\sigma\sigma}(\bfqp,z,z',\omega) v_{s1\sigma}(\bfqp,z',\omega) \:.
\end{equation}
The noninteracting response function is diagonal in the spin $\sigma$:
\begin{eqnarray} \label{L.3.2}
\chi_{s,\sigma\sigma'}(\bfqp,z,z',\omega) &=& \delta_{\sigma\sigma'}
\sum_{j=1}^{N_{\rm occ}} \sum_{l=1}^\infty F_{lj}(\bfqp,\omega)  \nonumber\\
&&
\times
\varphi_{j}(z)\varphi_{l}(z) \varphi_{j}(z') \varphi_{l}(z') \:,
\end{eqnarray}
where
\begin{eqnarray} \label{L.3.3}
F_{lj}(\bfqp,\omega) &=&  \int \! \frac{d^2 \kp}{(2\pi)^2} \!
\bigg[ \frac{\theta(\varepsilon_F - \varepsilon_j - \kp^2/2)}{\displaystyle \omega - \omega_{lj} - \bfqp\bfkp - \qp^2/2 + i \eta}
\nonumber\\
&& {}-
\frac{\theta(\varepsilon_F - \varepsilon_j - \kp^2/2)}{\displaystyle \omega + \omega_{lj} + \bfqp\bfkp + \qp^2/2 + i \eta}\bigg] .
\end{eqnarray}
Here, $\omega_{jk} = \varepsilon_{k} - \varepsilon_{j}$, and $\eta$ is a positive infinitesimal.
The linearized effective potential,  $v_{s1\sigma} = v_{1\sigma} + v_{\rm Hxc1\sigma}$,
consists of an external scalar perturbation plus a linearized Hartree-xc contribution:
\begin{eqnarray}\label{L.3.5}
v_{\rm Hxc1\sigma}(\bfqp,z,\omega) &=&  \sum_{\sigma'} \int dz'
\bigg[\frac{2\pi}{\qp}\: e^{-\qp|z-z'|} \\
&+&
f_{\rm xc,\sigma \sigma'}(\bfqp,z,z',\omega) \bigg]n_{1\sigma'}(\bfqp,z',\omega) \:. \nonumber
\end{eqnarray}
The xc kernel $f_{\rm xc,\sigma \sigma'}$ will be discussed in more detail below.

The following external perturbation triggers both sin\-gle-particle and collective excitations with a finite in-plane wave vector $\bfqp$:
\begin{equation}\label{L.3.7}
v_{1\sigma}(\bfqp,z,\omega) = S_\sigma^\pm E_0 e^{\qp z} \:,
\end{equation}
which couples to the charge $(+)$ and the  spin $(-)$ channel via
$S_\sigma^\pm = \delta_{\sigma,\uparrow} \pm \delta_{\sigma,\downarrow}$, respectively.
Having solved the response equation (\ref{L.3.1}) self-consistently, we obtain the absorption cross section as
\begin{equation}\label{L.3.9}
\sigma(\bfqp,\omega) = -\frac{2\omega}{E_0 \qp^2} \: \Im
\sum_\sigma S_\sigma^\pm \int dz\: e^{\qp z} n_{1\sigma}(\bfqp,z,\omega)\:.
\end{equation}
The absorption cross section, when plotted as a function of frequency, has peaks
at those frequencies that are resonant with an excitation energy of the system; the peak height is a measure of
the oscillator strength.

The alternative to calculating the absorption cross section
is to directly calculate the excitation energies of the system. The idea is
that an electronic excitation can be viewed as an electronic eigenmode, i.e., a dynamical response
of the system that is self-sustained and does not require an external perturbation. The characteristic
eigenmode frequencies are thus obtained as those frequencies $\Omega$ where the linear-response equation
has a nontrivial solution in the absence of an external perturbation.\cite{Petersilka1996,Ullrich2014}
The resulting general formalism for calculating excitation energies in TDDFT has the form of an eigenvalue
equation: \cite{Casida1995,Ullrich2012}
\begin{equation} \label{V.26}
\left( \begin{array}{cc} {\bf A} & {\bf K} \\ {\bf K} & {\bf A} \end{array} \right)
\left( \begin{array}{c} {\bf X} \\ {\bf Y} \end{array} \right)
= \Omega
\left( \begin{array}{cc} -{\bf 1} & {\bf 0} \\ {\bf 0} & {\bf 1} \end{array} \right)
\left( \begin{array}{c} {\bf X} \\ {\bf Y} \end{array} \right),
\end{equation}
where the matrix elements of $\bf A$ and $\bf K$ are given by
\begin{eqnarray} \label{V.27}
A_{ia \sigma,i'a'\sigma'}(\omega) &=& \delta _{ii'} \delta_{aa'} \delta_{\sigma\sigma'} \omega_{ai \sigma} +
K_{ia \sigma,i'a'\sigma'}(\omega)
\\
K_{ia \sigma,i'a'\sigma'}(\omega) &=&\int d^3r \int d^3r' \varphi_{i}^*(\bfr) \varphi_{a}(\bfr)
\bigg\{\frac{1}{|\bfr - \bfr'|}
\nonumber\\
&+& f_{\rm xc\sigma\sigma'}(\bfr,\bfr',\omega)\bigg\} \varphi_{i'}(\bfr')\varphi_{a'}^*(\bfr') \label{V.28}
\hspace{5mm}
\end{eqnarray}
and $i,i'$ and $a,a'$ run over occupied and unoccupied Kohn-Sham orbitals, respectively.
In almost all applications of this formalism one uses frequency-independent approximations for
the xc kernel.

Equation (\ref{V.26}) can be adapted in a rather straightforward manner to calculate inter- and intrasubband
charge and spin plasmon frequencies in quantum wells; all one needs to do is use the explicit form
$\varphi_j(\bfr) = A^{-1/2}\varphi_j(z)e^{i\bfkp \cdot \bfrp}$ of the single-particle wave functions and
then Fourier transform with respect to $\bfrp$.

Rather than giving the general formalism, let us consider the much simpler (but very important) quasi-2D case. Assume that only
the lowest subband is occupied, and consider the lowest intersubband plasmon modes at wavevector $\bfqp=0$. Ignoring the influence of
the third and higher subbands, the intersubband charge and spin plasmon frequencies are given by
\begin{equation}\label{20}
\Omega_{c,s}^2 = \omega_{21}^2 + \omega_{21} N_s (K_{\ua\ua} \pm K_{\ua\da})\:,
\end{equation}
where
\begin{eqnarray}\label{21}
K_{\sigma\sigma'} &=& \int dz \int dz' \varphi_1(z)\varphi_2(z)[-2\pi|z-z'|
\nonumber\\
&&{}+ f_{\rm xc,\sigma\sigma'}(z,z')]\varphi_1(z')\varphi_2(z').
\end{eqnarray}

\subsection{Exchange kernels}
The main purpose of this paper is to compare the performance of different approximate xc kernels
in the crossover from 3D to 2D. In the following we shall limit ourselves to the exchange-only case.
For systems that are not spin polarized, the spin-resolved exchange kernel $f_{\rm x,\sigma\sigma'}$
is obtained from the spin-unresolved exchange kernel $f_{\rm x}$ as
\begin{equation}\label{A2}
f_{\rm x,\sigma\sigma'} =2 \delta_{\sigma\sigma'} f_{\rm x}\:.
\end{equation}
We compare three different frequency-independent exchange kernels: ALDA, PBE, and PGG.
The ALDA exchange kernel is defined as follows:
\begin{equation}
f_{\rm x}^{\rm ALDA}(\bfr,\bfr') = \left. \frac{d^2 e_{\rm x}^h(\bar n)}{d \bar n ^2}\right|_{\bar n = n(\bfr)}
\delta(\bfr-\bfr') \:,
\label{eqn:ALDAxc}
\end{equation}
where $e_{\rm x}^h(n)$ is the exchange energy density of a homogeneous electron liquid of density $n$.\cite{GV}
Hence, the 3D and 2D ALDA exchange kernels are given by
\begin{eqnarray}
f_{\rm x,3D}^{\rm ALDA}(\bfr,\bfr') &=& -[9\pi n^2(\bfr)]^{-1/3}\delta(\bfr-\bfr')
\\
f_{\rm x,2D}^{\rm ALDA}(\bfrp,\bfrp') &=& - [ \pi n_{\rm 2D}(\bfrp)/2]^{-1/2} \delta(\bfrp-\bfrp') \:.
\end{eqnarray}

The PBE functional\cite{PBE1996} is probably the most widely used GGA; it is defined only for 3D systems.
The explicit expression for the PBE exchange kernel turns out to be quite lengthy, and is given
in Appendix A.

In contrast with ALDA and PBE, the so-called PGG functional\cite{Petersilka1996,Petersilka1998} is a
nonlocal orbital functional, given by
\begin{equation}\label{PGGkernelr}
f_{\rm x}\PGG(\bfr,\bfr') = - 2
\frac{\left|\sum_{j=1}^{N_{\rm occ}} \varphi_{j}(\bfr) \varphi^*_{j}(\bfr')\right|^2}
{|\bfr - \bfr'| n(\bfr) n(\bfr')} \:,
\end{equation}
where the sum runs over $N_{\rm occ}$ doubly occupied orbitals.
PGG can be viewed as an approximation to the exact exchange kernel.\cite{Ullrich2012}
We give the explicit form of the PGG kernel for quasi-2D systems and for the 2D limit in Appendix B, and
discuss its relation to exchange-only ISTLS in Appendix C.

\section{Results and Discussions}

\begin{figure*}
\begin{center}
\includegraphics[angle=0, width=\textwidth]{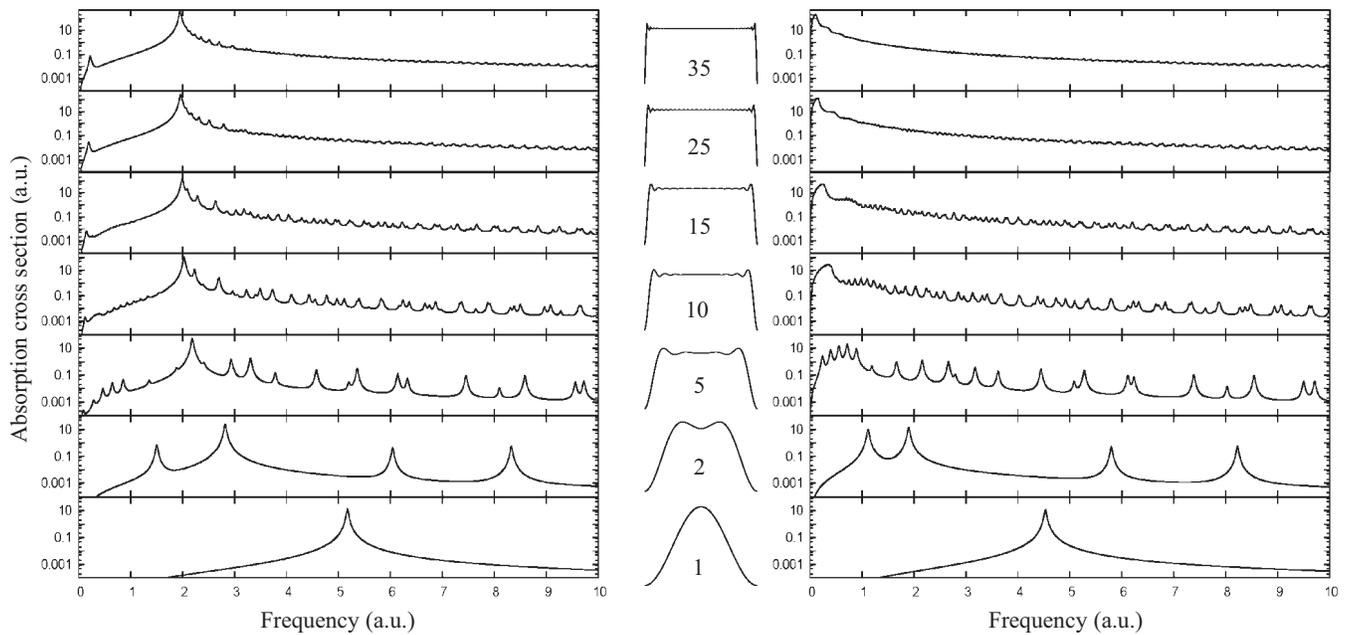}
\end{center}
\caption{Photoabsorption cross section for $\qp=0$ intersubband excitations in quantum wells. Left panels: charge-density excitations.
Right panels: spin-density excitations. Insets: density profiles at given values of $N_{\rm occ}$. The calculations were
done with the 3D ALDA exchange kernel.}
\label{fig2}
\end{figure*}

\subsection{Plasmons: from bulk to quasi-2D}

Plasmons in homogeneous electron liquids have been thoroughly studied for many decades.\cite{March1995}
The plasmon dispersions in 2D and 3D follow from the exact conditions
\begin{eqnarray} \label{3Dplasmons}
\left[\frac{4\pi}{q^2} + f_{\rm xc,3D}(q,\Omega_{\rm 3D})\right]\chi_0^{\rm 3D}(q,\Omega_{\rm 3D}) &=& 1\\
\left[\frac{2\pi}{\qp} + f_{\rm xc,2D}(\qp,\Omega_{\rm 2D})\right] \chi_0^{\rm 2D}(\qp,\Omega_{\rm 2D})&=& 1 \:, \label{2Dplasmons}
\end{eqnarray}
where $\chi_0^{\rm 3D}(q,\Omega)$ and $\chi_0^{\rm 2D}(\qp,\Omega)$ are the 3D and 2D Lindhard functions.\cite{GV}
In the limit of small wavevectors, one obtains
\begin{equation}
\Omega_{\rm 3D}(q\to 0) = \omega_{\rm pl}\left[ 1 + \left( \frac{3 (k_{F}^{\rm 3D})^2}{10\omega_{\rm pl}^2} +
\frac{f_{\rm xc,3D}(0,\omega_{\rm pl})}{8\pi}\right)q^2\right],
\end{equation}
where $\omega_{\rm pl} = \sqrt{4\pi n}$ is the classical plasma frequency of a 3D electron liquid of density $n$,
and $k_F^{\rm 3D}$ is the associated Fermi wavevector. The corresponding relation in 2D is
\begin{equation}
\Omega_{\rm 2D}(\qp \to 0) = k_{F}^{\rm 2D}\sqrt{\qp}\left[ 1 + \frac{\qp}{2\pi} f_{\rm xc,2D}(0,0)\right]^{1/2}.
\end{equation}
$\Omega_{\rm 3D}(q)$ and $\Omega_{\rm 2D}(\qp)$ both describe charge plasmons (i.e., collective oscillations of
the charge density $n$). There are no corresponding 3D and 2D spin plasmons (i.e., collective oscillations
of the spin density) as long as the system is not magnetic: the reason is that the 3D and 2D spin plasmons
fall into the respective particle-hole continua and are hence Landau damped.

Suppose now that we start from a homogeneous 3D system and let one of its dimensions, say $z$, become confined:
this defines a neutral jellium slab.\cite{Dobson1992,Schaich1994}
Let us  consider a jellium slab that corresponds to the
quantum well model with hard boundaries that we described in Section II.A.
What happens to the plasmon mode as the width $L$ of this system shrinks down to the quantum limit?

As soon as  $L$  becomes finite, the collective excitations are described using the formalism
of intersubband plasmons. We consider the case where the average 3D density $\bar n$ in the well is constant, letting
\begin{equation}
\bar n = N_s/L \:.
\end{equation}
If $L$ is very large, the difference between two consecutive energy levels $\varepsilon_j$ and $\varepsilon_{j+1}$,
see Eq. (\ref{II.28}), is very small, and a large number of subbands is occupied. As $L$ shrinks, the level
spacing increases and fewer and fewer subbands are occupied. Let $L_\nu$ be that width where the Fermi energy $\varepsilon_F$
coincides with the $\nu$th level $\varepsilon_\nu$. From Eqs. (\ref{II.28}) and (\ref{II.31}) it is straightforward to
show that
\begin{equation}
L_\nu^3 = \frac{\pi \nu}{12\bar n}( 4\nu^2 - 3\nu - 1 ) \:,
\end{equation}
where we used $\sum_{j=1}^\nu j^2 = \nu(\nu+1)(2\nu+1)/6$. In particular, for $\nu=2$ we have
\begin{equation}\label{II.42}
L_2 = \left(\frac{3\pi}{2\bar n}\right)^{1/3} \:.
\end{equation}
For $L<L_2$, only the lowest subband is occupied (the quantum limit).  Equation (\ref{II.42}) can also be rewritten
in terms of the 2D Wigner-Seitz radius $r_s^{\rm 2D}$ as\cite{Pollack2000}
\begin{equation} \label{II.43}
L_2 = \sqrt{\frac{3\pi}{2N_s}} =\pi r_s^{\rm 2D} \sqrt{\frac{3}{2}} \approx 3.85 r_s^{\rm 2D} \:.
\end{equation}

Figure \ref{fig2} shows ALDA intersubband excitation spectra at $\qp=0$, in the charge and spin channel, for quantum wells
with different numbers of occupied subbands, ranging from $N_{\rm occ}=1$ to 35. $L$ and $N_s$ are chosen such that the
average density remains constant at $\bar n = 0.30 \: {a_0^*}^{-3}$.
The insets in the middle show how the density profile becomes more and more square shaped as $N_{\rm occ}$ increases.

In the quasi-2D limit where $N_{\rm occ}=1$, the spectra only show a single peak in the energy range below 10 a.u.: the
intersubband charge plasmon at 5.17 a.u. (left bottom panel) and spin plasmon at 4.53 a.u. (right bottom panel).
As more subbands become
occupied, the spectra acquire more and more peaks, and eventually approach very simple limits for large $N_{\rm occ}$.

At $N_{\rm occ}=35$, the charge-density excitation spectrum is dominated by a single peak at 1.94 a.u., which is
the bulk plasmon frequency $\omega_{\rm pl}$ corresponding to $\bar n$. There is also a small peak around 0.22 a.u.,
which corresponds to the surface plasmon of a large jellium slab with a sharp density profile.\cite{Liebsch1997}
On the other hand, the spin-density excitation spectrum has become essentially featureless; in other words, the
spin plasmon has disappeared, as expected.

Thus, there is a seamless transition between the 3D bulk plasmon and the intersubband plasmons as the 2D limit is approached.
In this regime, the 3D ALDA (or any 3D semilocal functional) is appropriate.

\subsection{2D Limit of intersubband plasmons}

\begin{figure}
\begin{center}
\includegraphics[angle=0, width=8.5cm]{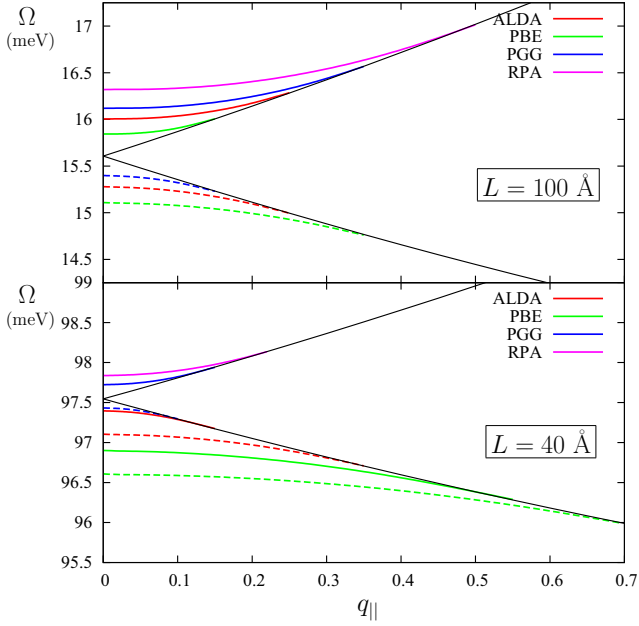}
\end{center}
\caption{(Color online) Intersubband plasmon dispersions $\Omega(\qp)$, for $N_s=10^{12} \: \rm cm^{-2}$
and well widths 100 {\AA} and 40 {\AA}. The black full lines indicate the intersubband p-h continuum.
The RPA only
gives intersubband charge plasmons; ALDA, PBE and PGG give both charge (full lines) and spin plasmons (dashed lines).
ALDA and PBE break down when their charge plasmons falls below the p-h continuum. }
\label{fig3}
\end{figure}

We now focus on the situation where only the first subband is occupied ($N_{\rm occ}=1$), i.e., we consider quantum wells
of width $L <L_2$. Figure \ref{fig3} shows the intersubband charge and spin plasmon dispersions
for quantum wells with $N_s = 10^{12}\:\rm cm^{-2}$ and $L=100$ and 40 {\AA}, respectively, calculated with RPA, ALDA, PBE and PGG.
In all cases, the charge plasmon dispersion lies above the spin plasmon dispersion (except for RPA, which has no spin plasmon).
However, the position of the intersubband plasmon dispersions relative to the particle-hole (p-h) continuum varies.

For the 100 {\AA} wide quantum well we find that the charge plasmon branches are above the p-h continuum and spin plasmon
branches are below. For the 40 {\AA} well, however, the charge plasmon branch has moved {\em below}\cite{DasSarma1993}
the p-h continuum for ALDA and PBE, but not for RPA and PGG. This is a remarkable difference between semilocal
and orbital-dependent exchange functionals, and we will now investigate this effect in more detail.

\begin{figure}
\begin{center}
\includegraphics[angle=0, width=8.5cm]{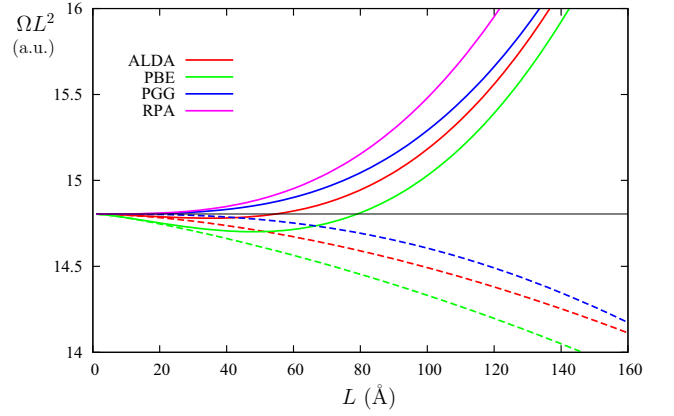}
\end{center}
\caption{(Color online) Intersubband plasmon energies, at $\qp=0$,  versus well width $L$, for $N_s=10^{12} \: \rm cm^{-2}$.
The horizontal line indicates the lowest p-h transition $\omega_{21}$ (all energies are scaled by $L^2$). The RPA only
gives intersubband charge plasmons; ALDA, PBE and PGG give both charge (full lines) and spin plasmons (dashed lines). ALDA and PBE break down when
the charge plasmon falls below the p-h line.}
\label{fig4}
\end{figure}

Let us consider the case $\qp=0$ and keep the sheet density $N_s$
fixed. As $L\to 0$, the system transitions from quasi-2D to strictly 2D.\cite{2Dfootnote}
In this limit, the intersubband excitation energies become infinitely large, because the system is so strongly confined
in the plane that density fluctuations perpendicular to the quantum well plane (see Fig. \ref{fig1}) become impossible.
However, it is interesting to observe how the intersubband plasmons behave as this limit is approached. This is
shown in Fig. \ref{fig4}.

We have calculated the $\qp=0$ intersubband charge and spin plasmon
energies with RPA (charge plasmon only), ALDA, PBE, and PGG. According to Eq. (\ref{II.28})
the lowest p-h transition energy is $\omega_{21} = 3\pi^2/2L^2$. Hence, $\omega_{21}L^2$ is constant,
as indicated by the thin horizontal line in Fig. \ref{fig4}.
As $L$ becomes smaller, the plasmon energies (scaled by $L^2$)
approach and eventually merge with the p-h line.

The RPA plasmon energy follows from Eq. (\ref{20}) as
\begin{equation}
(\Omega_c^{\rm RPA} L^2)^2 = \frac{9\pi^4}{4} +  \frac{20 \pi N_s L^3}{3} \:,
\end{equation}
where the Hartree part of the intersubband matrix element (\ref{21}) is given by
\begin{equation}
-2\pi\int dz \int dz' \varphi_1(z) \varphi_2(z) |z-z'| \varphi_1(z')\varphi_2(z') = \frac{20L}{9\pi}. \label{38}
\end{equation}
Hence, the RPA charge plasmons are always shifted {\em above} the p-h line, but the separation vanishes as $L\to 0$.

In ALDA, we find
\begin{eqnarray}\label{36}
(\Omega_c^{\rm ALDA} L^2)^2 &=& \frac{9\pi^4}{4} +  \frac{20 \pi N_s L^3}{3}
- c_1 \left(48 \pi^2 N_s L^5\right)^{1/3}  \\
(\Omega_s^{\rm ALDA} L^2)^2 &=& \frac{9\pi^4}{4}
-  c_1 \left(48 \pi^2 N_s L^5\right)^{1/3},
\end{eqnarray}
where $c_1=\int_0^\pi dx \sin^2(2x)\sin^{2/3}(x) = 1.20027$. For the PBE
and PGG plasmon energies no simple analytic expressions exist; however, numerical evaluation is straightforward
using the formulas in the Appendix.

As can be seen from Fig. \ref{fig4}, the ALDA and PBE charge plasmons cross over the p-h line: this
happens at $L=54.6$ {\AA} in ALDA and at $L=79$ {\AA} in PBE. No such crossover is observed for PGG.

\begin{figure}
\begin{center}
\includegraphics[angle=0, width=8.5cm]{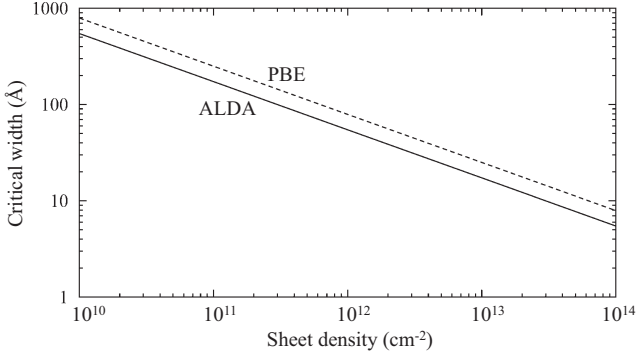}
\end{center}
\caption{ Critical width $L_{\rm crit}^{\rm inter}$ at which the intersubband plasmon breakdown occurs,
as a function of sheet density $N_s$. Full line: ALDA, dashed line: PBE.}
\label{fig5}
\end{figure}

The critical width $L_{\rm crit}^{\rm inter}$ at which the crossover occurs in ALDA and PBE is plotted in Fig. \ref{fig5}
as a function of the sheet density $N_s$.
In ALDA we can use Eq. (\ref{36}) to find the analytical result
\begin{equation}
L_{\rm crit}^{\rm inter} = \frac{3c_1^{3/4}}{5\sqrt{N_s}}\left(\frac{5}{4\pi}\right)^{1/4} = \frac{0.546}{\sqrt{N_s}}
\: \mbox{a.u.}
\end{equation}
For PBE, we obtain numerically $L_{\rm crit}^{\rm inter} = 0.79/\sqrt{N_s}$ a.u.
In terms of the 2D Wigner-Seitz radius, this becomes $L_{\rm crit}^{\rm inter} = 0.975\, r_s^{\rm 2D}$ and
$1.40 \, r_s^{\rm 2D}$ for ALDA and PBE, respectively. In the case of ALDA, this is about 4 times smaller than $L_2$ [Eq. (\ref{II.43})],
the width of the quantum well below which only the lowest subband is occupied; in the case of PBE, it is about 3 times smaller.

In PGG, we find that the charge and spin plasmons always lie above and below the p-h continuum, respectively.
This is similar to the case of excitation energies in atoms, where the bare Kohn-Sham exictations are
found to lie between the singlet and triplet excitations.\cite{Savin1998,Burke2002,Baerends2013}
Hence, the crossover of ALDA and PBE indicates a general failure of semilocal functionals in the 2D limit
of intersubband transitions.

However, it is important to note that this failure does not appear to be a catastrophic breakdown, as
in the case of the diverging exchange energy that we discussed in the Introduction. The intersubband plasmons may
have a wrong position with respect to the p-h continuum, but they still exist as collective modes, and deviate
not too far from the PGG results. Furthermore, the separation between charge and spin plasmons (the analog of
the singlet-triplet splitting in atoms) remains well described in ALDA and PBE for all $L$.

In practice, the width of quantum wells is limited by the underlying material
(for GaAs, the lattice constant is 5.65 \AA). Typical semiconductor quantum wells have widths of several hundreds of \AA,
so that one is usually sufficiently far away from the critical widths where the ALDA breaks down for the intersubband dynamics,
except for situations where $N_s$ is very small.

\subsection{2D limit of intrasubband plasmons}

\begin{figure}
\begin{center}
\includegraphics[angle=0, width=\columnwidth]{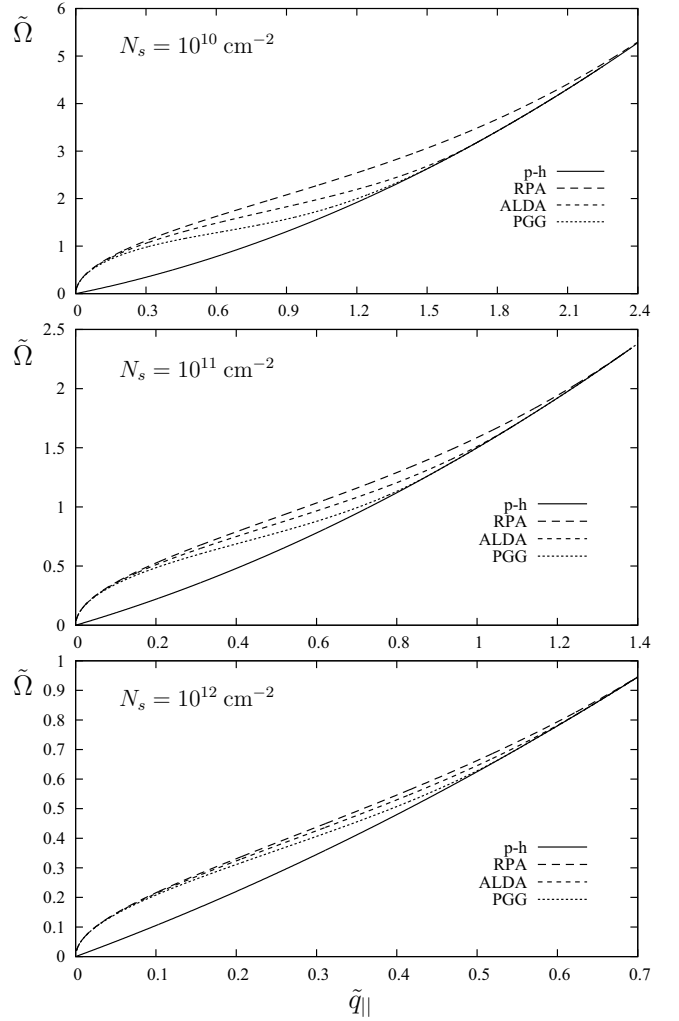}
\end{center}
\caption{Plasmon dispersions $\Omega(\qp)$ for strictly 2D systems with sheet densities $N_s=10^{10}$, $10^{11}$,
and $10^{12}\: \rm cm^{-2}$, calculated with RPA, 2D ALDA and PGG. The full lines denote the upper boundaries of the particle-hole (p-h) continuum.
Here, $\tilde\qp = \qp/k_F^{\rm 2D}$ and $\tilde \Omega = \Omega/(k_F^{\rm 2D})^2$.}
\label{fig6}
\end{figure}

Let us now consider the intrasubband plasmons in a quantum well with $N_{\rm occ}=1$, in the limit where $L\to 0$.
For convenience, we shift the bottom of the quantum well potential such that the lowest subband level
$\varepsilon_1=0$. Assuming, furthermore, that the second and higher subband levels are energetically well separated from
the lowest subband, the response function (\ref{L.3.2}) is given by
\begin{equation}
\chi_{s,\sigma\sigma'}(\bfkp,z,z',\omega) = \delta_{\sigma\sigma'}\Phi(z,z')\chi_0^{\rm 2D}(\kp,\omega)\:,
\end{equation}
where $\chi_0^{\rm 2D}(\kp,\omega)$ is the 2D Lindhard function, and where we abbreviate $\Phi(z,z') = \varphi_1^2(z)\varphi_1^2(z')$.
The response equation (\ref{L.3.1}) for the eigenmodes then becomes
\begin{eqnarray}
n_1(\qp,z',\Omega) &=& \int dz_1 \Phi(z',z_1) \chi_0^{\rm 2D}(\qp,\Omega)
\nonumber\\
&\times&
\int dz_2 f_{\rm Hxc}(\qp,z_1,z_2) n_1(\qp,z_2,\Omega) \:. \quad
\end{eqnarray}
Multiply both sides with $\varphi_1^2(z)f_{\rm Hxc}(\qp,z,z')$ and integrate over $z$ and $z'$. Then, $n_1$ cancels out and
we are left with the condition
\begin{eqnarray} \label{quasi2Dplasmon}
1 &=& \int dz \! \int dz'  \Phi(z,z') \! \left[\frac{2\pi}{\qp}e^{-\qp|z-z'|} + f_{\rm xc}(z,z')\right] \nonumber\\
&&{}\times
\chi_0^{\rm 2D}(\qp,\Omega)\:.
\end{eqnarray}
The intrasubband plasmons of the quasi-2D quantum well are those frequencies $\Omega$ where Eq. (\ref{quasi2Dplasmon})
is satisfied.
The question is now this: if $L\to 0$, will Eq. (\ref{quasi2Dplasmon}) turn into Eq. (\ref{2Dplasmons}) for the 2D plasmons?

A straightforward calculation shows that this is indeed the case for the Hartree part, as expected.
Using the particle-in-a-box wave function (\ref{box}) we obtain
\begin{eqnarray}
\lefteqn{
\int dz \! \int dz' \: \Phi(z,z')e^{-\qp|z-z'|}
= \frac{\qp L}{(\qp^2L^2 + 4\pi^2)^2}
}
\nonumber\\
&&
\times
\left\{3\qp^2 L^2 + 20 \pi^2 +
\frac{32\pi^4}{\qp^3L^3}(e^{-\qp L}-1 + \qp L)\right\}\nonumber\\
&&\longrightarrow 1 \quad \mbox{for} \quad L\to 0.
\end{eqnarray}
For the PGG exchange kernel, it is straightforward to show that
\begin{equation}
\int dz \! \int dz' \: \Phi(z,z') f_{\rm x}^{\rm PGG}(\qp,z,z')
\longrightarrow f_{\rm x,\rm 2D}^{\rm PGG}(\qp)
\end{equation}
for $L\to 0$, where $f_{\rm x}^{\rm PGG}(\qp,z,z')$ and $f_{\rm x,\rm 2D}^{\rm PGG}(\qp)$ are given in Appendix B,
see Eqs. (\ref{PGGkernel} and (\ref{PGGkernel2D}). Thus, the PGG exchange kernel behaves correctly
in the 2D limit.

However, it is hardly surprising to find that the ALDA does not give the correct 2D limit. We have
\begin{equation}
\int dz \! \int dz' \: \Phi(z,z') f_{\rm x,3D}^{\rm ALDA}(z,z') =
-\frac{2c_2}{3\pi} \left( \frac{6}{\pi L}\right)^{1/3} n_{\rm 2D}^{-2/3},
\end{equation}
where $c_2 = \int_0^{\pi} dx \sin^{8/3}(x) = 1.4003$. This clearly disagrees with the form of
$f_{\rm x,2D}^{\rm ALDA}=-\sqrt{2/\pi n_{\rm 2D}}$, and in fact diverges as $L\to 0$.
Other semilocal functionals such as PBE show similar trends.

Figure \ref{fig6} shows the plasmon dispersions in the strictly 2D limit, calculated by solving Eq. (\ref{2Dplasmons}).
The ALDA and PGG calculations were done with the 2D exchange kernels $f_{\rm x,2D}^{\rm ALDA}$
and $f_{\rm x,2D}^{\rm PGG}$, respectively.
The upper boundary of the particle-hole continuum is given by the relation
$\Omega_{p-h} = \tilde \qp^2/2 + \tilde \qp$, where $\tilde\qp = \qp/k_F^{\rm 2D}$.
One observes that the RPA plasmon dispersion always lies above ALDA and PGG, reflecting the downshift of
excitation energies caused by exchange.

\begin{figure}
\begin{center}
\includegraphics[angle=0, width=\columnwidth]{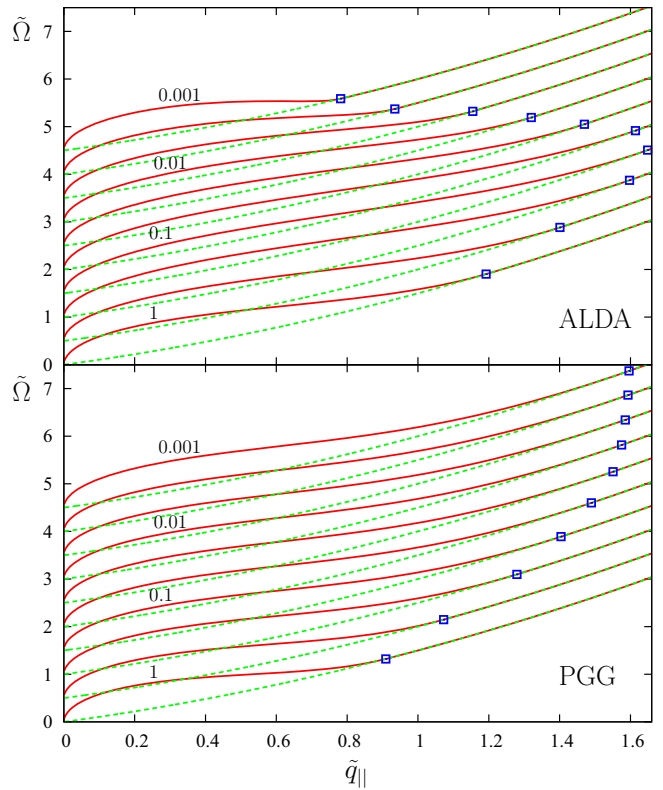}
\end{center}
\caption{(Color online) Intrasubband plasmon dispersions for quantum wells with sheet density $N_s=10^{10}\: \rm cm^{-2}$,
for different widths $L=\lambda L_2$, where $\lambda$ takes on the values 1, 0.5, 0.2, 0.1, 0.05, 0.02, 0.01,
0.005, 0.002, and 0.001. $L_2=217$ nm is the largest width for which only the lowest subband is occupied.
The individual plasmon dispersions are offset for clarity. The dashed lines are the upper boundaries of the p-h continuum.
The squares indicate the wavevector $\tilde q_{|| p-h}$ where the plasmons enter the p-h continuum.
Top panel: 3D ALDA. Bottom panel: PGG.}
\label{fig7}
\end{figure}

Figure \ref{fig7} compares the intrasubband plasmon dispersions of PGG and 3D ALDA for well widths
$L= \lambda L_2$, where we let the scaling parameter $\lambda$ take on values between 1 and 0.001
[recall that $L_2$, Eq. (\ref{II.42}), is the maximum well
width for which only the lowest subband is occupied for a given $N_s$]. The sheet density
is $N_s=10^{10}\: \rm cm^{-2}$, and we have $L_2 = 217$ nm.

As expected, PGG nicely approaches the 2D limit that was shown in Fig. \ref{fig6}. For $\lambda<0.01$,
the intrasubband plasmon dispersion becomes indistinguishable from the strictly 2D limit.

The situation is drastically different for the ALDA. As $\lambda$ decreases from 1 to 0.1, the intrasubband
dispersion appears to approach the 2D limit. However, below $\lambda=0.1$ the 3D ALDA breaks down, and
the intrasubband plasmon dispersion becomes more and more suppressed, that is, it begins to merge with the p-h continuum
at smaller and smaller wavevectors. As $\lambda\to 0$, the intrasubband plasmon completely
disappears, rather than approaching the 2D plasmon shown in Fig. \ref{fig6}.

\begin{figure}
\begin{center}
\includegraphics[angle=0, width=\columnwidth]{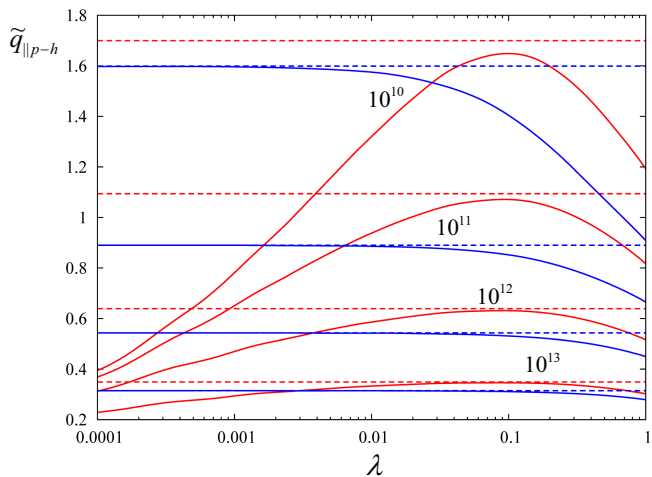}
\end{center}
\caption{(Color online) Wavevector $\tilde q_{|| p-h}$ at which the intrasubband plasmon merges with the p-h continuum,
plotted versus well width scaling factor $\lambda$, calculated with PGG (blue) and ALDA (red). The dashed lines indicate
the respective limits for the strictly 2D case. The calculations were done for sheet densities
$N_s = 10^{10}, 10^{11}, 10^{12}, \:\mbox{and}\:  10^{13} \: \rm cm^{-2}$, as indicated. The breakdown of the 3D ALDA
occurs around $\lambda=0.1$ for all $N_s$.}
\label{fig8}
\end{figure}

We have repeated these calculations for several different values of the sheet density $N_s$, focusing on
the wavevector $\tilde q_{|| p-h}$ where the intrasubband plasmon enters the p-h continuum, as indicated by
the blue squares in Fig. \ref{fig7}.

Figure \ref{fig8} shows $\tilde q_{|| p-h}$ versus the well width scaling factor $\lambda$
for $N_s = 10^{10}, 10^{11}, 10^{12}, \:\mbox{and}\:  10^{13} \: \rm cm^{-2}$, calculated with ALDA and PGG.
For PGG we see in each case that $\tilde q_{|| p-h}$ smoothly approaches its limiting value for the strictly 2D plasmon,
shown by the dashed line. The ALDA initially approaches the 2D limit as $\lambda$ decreases from 1.
However, around $\lambda=0.1$ all ALDA curves turn around and rapidly drop off, moving away from the 2D limit.

Thus, we find that the 3D ALDA exchange kernel behaves reasonably as long as the well width is sufficiently large.
The breakdown for intrasubband (in-plane) dynamics occurs for $L_{\rm crit}^{\rm intra}\approx 0.1L_2 \approx 0.4 r_s^{\rm 2D}$.
Interestingly, this is significantly smaller than the critical {\em inter}subband width $L_{\rm crit}^{\rm inter}\approx r_s^{\rm 2D}$,
see Section III.B.

\section{Conclusions}

In this paper we have carried out systematic numerical studies of the electron dynamics in quantum wells
whose width $L$ crosses over from the 3D to the quasi-2D regime (where only the lowest subband is occupied, but the
finite size is still relevant) and finally to the strictly 2D limit (where $L=0$).  The purpose was a comparison
of different classes of exchange kernels in TDDFT: standard semilocal kernels (such as ALDA and PBE) and nonlocal
kernels (such as PGG and ISTLS). ALDA and PBE are based on the electron gas as reference system, whereas PGG and ISTLS
are orbital functionals, whose definition does not invoke any reference system.

The main conclusion does not come as a surprise: ALDA and PBE fail in the 3D-2D crossover, PGG succeeds.
This is already well known for the ground state,\cite{Pollack2000,Kim2000,Constantin2008a,Constantin2008b}
and there was no reason to expect otherwise for the dynamical case.
However, the details are interesting and of practical relevance.

First of all, we discover a universal behavior of the breakdown of the inter- and intrasubband dynamics
in 3D ALDA. At a critical well width of $L_{\rm crit}^{\rm inter}\approx r_s^{\rm 2D}$, intersubband plasmons
are no longer qualitatively correctly described (the charge plasmon falls below the single-particle excitation $\omega_{21}$).
For well widths below $L_{\rm crit}^{\rm intra} \approx 0.4 r_s^{\rm 2D}$, intrasubband plasmon dispersions
start to become suppressed compared to the 2D limit. The interesting finding is thus that
$L_{\rm crit}^{\rm intra} < L_{\rm crit}^{\rm inter}$,  so the in-plane dynamics
appears to be well described using the 3D ALDA down to much smaller widths than the out-of-plane dynamics.

Compared to the ground state, the failure of the (semi)local xc functionals in the dynamical case is of a different nature.
In fact, while the exchange energy diverges for $L\to 0$, intersubband plasmons can still be reasonably described
(apart from the fact that they drop below the p-h continuum,\cite{DasSarma1993} which is an artifact of these functionals).
In turn, intrasubband plasmon dispersions become suppressed and cease to exist, instead of approaching the limit
of 2D plasmons.

In practice, it is important to know for what quantum well widths the 3D ALDA is still applicable.
For instance, if $N_s=10^{11} \: \rm cm^{-2}$ (which is a very typical value for many semiconductor quantum well samples),
we find $L_{\rm crit}^{\rm inter}= 17$ nm for GaAs, which is rather narrow. Higher sheet densities allow one to push this limit
to even narrower wells; and the breakdown for intrasubband dynamics occurs at even smaller well widths, as low as a few \AA.
This is certainly good news, considering the popularity of the ALDA and its ease of implementation.
We also find that these values can be significantly higher for the PBE; in other words, using gradient-corrected
xc functionals for quantum wells does not seem to pay off.

Clearly, the best option to describe the dynamics in strongly confined systems is using nonlocal orbital functionals
such as PGG or ISTLS, since these are not tied to a particular choice of reference system (such as the 2D or 3D ALDA)
and hence have no problem with dimensional crossover.

Finally, let us say a few words about correlation. In the ground-state,\cite{Pollack2000,Kim2000,Constantin2008a,Constantin2008b}
it was observed that local and semilocal correlation functionals break down in a similar manner as exchange functionals.
This will also be the case for the dynamics. However, nonlocal, orbital-dependent correlation functionals
are much more complicated than exchange functionals; for instance, implementing the ISTLS beyond exchange in linear response
will remain a task for the future.

There is another aspect of correlation that is unique to the dynamical case, namely, it leads to dissipation of
plasmon excitations even outside the particle-hole continuum. Plasmon damping in quantum wells has been
studied within time-dependent current-DFT,\cite{Ullrich1998a,Ullrich1998b,Ullrich2001,Ullrich2002}
using the complex and frequency-dependent xc kernel of
Vignale and Kohn.\cite{Vignale1996,Vignale1997} This xc kernel is a local approximation of the current,
and can lead to overdamping of charge plasmons.\cite{Ullrich1998b,DAgosta2007}
The effect is even more dramatic for spin plasmons, where the damping due to the spin Coulomb drag effect
is significantly overestimated using a local approximation.\cite{DAmico2006}
Again, it is found that the cure to this overdamping is provided by orbital functionals.\cite{DAmico2013}

\begin{acknowledgments}
This work was supported by DOE Grant No. DE-FG02-05ER46213.
\end{acknowledgments}

\appendix

\section{The PBE exchange kernel}
\subsection{PBE exchange energy}
The PBE exchange energy functional is defined as\cite{PBE1996}
\begin{equation} \label{A1}
E_{\rm x}^{\rm PBE}[n] = \int d^3r' \: e_{\rm x}^h(n) \left[1 + \kappa - \frac{\kappa}{1 + \mu s^2/\kappa}\right].
\end{equation}
Here, the exchange energy density of a homogeneous 3D electron liquid  of density $n$ is
\begin{equation}
e_{\rm x}^h(n) = -\frac{3c}{4} \, n^{4/3} \:, \qquad
c =  \left( \frac{3}{\pi}\right)^{1/3}.
\end{equation}
In Eq. (\ref{A1}), $\kappa=0.804$ and $\mu=0.21951$ are parameters given in atomic units. The quantity $s$ is defined as
$s = |\nabla n|/2nk_F^{\rm 3D}$, where $k_F^{\rm 3D}=(3\pi^2 n)^{1/3}$ is the Fermi wavevector. Thus,
\begin{equation}
s = \frac{|\nabla n|}{2(3\pi^2)^{1/3}n^{4/3}}\:.
\end{equation}
Putting this into Eq. (\ref{A1}), we obtain
\begin{equation}
E_{\rm x}^{\rm PBE}[n] = \int d^3r' \: e_{\rm x}^h(n) \left[1 + \kappa - \frac{\kappa}{1 + \gamma |\nabla n|^2/n^{8/3}}\right],
\end{equation}
where
$\gamma = (\mu/4\kappa) (3\pi^2)^{-2/3}=0.007132$ a.u. For what follows, it is convenient to introduce the
abbreviation
\begin{equation}
g(\bfr) = 1 + \gamma |\nabla n(\bfr)|^2/n(\bfr)^{8/3} \:.
\end{equation}

\subsection{PBE exchange potential}
The PBE exchange potential it its spin-unresolved form is given by
\begin{eqnarray}
v_{\rm x}^{\rm PBE}(\bfr) &=& \frac{\delta E_{\rm x}^{\rm PBE}[n]}{\delta n(\bfr)}
\nonumber\\
&=&
\int d^3r'  \left( \frac{\delta e_{\rm x}^h(n(\bfr'))}{\delta n(\bfr)}\right)\!\left[1 + \kappa - \frac{\kappa}{g(\bfr')}\right]
\nonumber\\
&-&
\int d^3r' e_{\rm x}^h(n(\bfr')) \frac{\delta}{\delta n(\bfr)}\! \left(\frac{\kappa}{g(\bfr')}\right).
\end{eqnarray}
The first part is easy, with
\begin{displaymath}
\frac{\delta e_{\rm x}^h(n(\bfr'))}{\delta n(\bfr)}=-c n(\bfr')^{1/3}\delta(\bfr' - \bfr) \:.
\end{displaymath}
The second part requires more effort, involving functional derivatives of the gradient of $n$, which leads to gradients of delta
functions.  The final result is
\begin{eqnarray}
v_{\rm x}^{\rm PBE}(\bfr)
&=&
-cn(\bfr)^{1/3} \left[1 + \kappa - \frac{\kappa}{g(\bfr)}\right]
\nonumber\\[2mm]
&+&
\frac{3c}{4}  n(\bfr)^{-4/3}\:
\nabla \left[\frac{2\kappa\gamma}{g(\bfr)^2} \: \right]\cdot \nabla n(\bfr)
\nonumber\\[2mm]
&+&
\frac{3c}{4}n(\bfr)^{-4/3}\:
\frac{2\kappa\gamma}{g(\bfr)^2}\:\nabla^2 n(\bfr) \:.
\end{eqnarray}
The spin-dependent version of the PBE exchange energy functional follows from the spin-scaling relation
\begin{equation}
E_{\rm x}[n_\ua,n_\da] = \frac{1}{2} E_{\rm x}[2n_\ua] + \frac{1}{2} E_{\rm x}[2n_\da] \:.
\end{equation}
This gives the spin-resolved exchange potential
\begin{equation}
v_{\rm x \sigma}^{\rm PBE}(\bfr) = v_{\rm x}^{\rm PBE}[2n_\sigma](\bfr) \:.
\end{equation}
For a system whose density is not spin polarized we have $n_\ua = n_\da = n/2$. In this case, all potentials
are the same, i.e., $v_{\rm x \ua}^{\rm PBE}(\bfr)=v_{\rm x \da}^{\rm PBE}(\bfr)=v_{\rm x}^{\rm PBE}(\bfr)$.

\subsection{PBE exchange kernel}

The parallel-spin exchange kernel is defined as follows:
\begin{equation}
f_{\rm x,\sigma \sigma}^{\rm PBE}(\bfr,\bfr') = \frac{\delta v_{\rm x\sigma}^{\rm PBE}(\bfr)}{\delta n_{\sigma}(\bfr')}
\end{equation}
(in the exchange-only case, the antiparallel-spin kernel is  zero).
For spin-unpolarized systems, we have
\begin{equation}
f_{\rm x,\ua\ua}^{\rm PBE}(\bfr,\bfr') = f_{\rm x,\da\da}^{\rm PBE}(\bfr,\bfr')
= 2\:f_{\rm x}^{\rm PBE}(\bfr,\bfr'),
\end{equation}
where
\begin{equation}
f_{\rm x}^{\rm PBE}(\bfr,\bfr') = \frac{\delta v_{\rm x}^{\rm PBE}[n](\bfr)}{\delta n(\bfr')} \:.
\end{equation}
After a rather lengthy calculation, one obtains
\begin{eqnarray}
f_{\rm x}^{\rm PBE}(\bfr,\bfr')
&=&
-\frac{c}{3}n(\bfr)^{-2/3}\delta(\bfr - \bfr') \left[1 + \kappa - \frac{\kappa}{g(\bfr)}\right]
\nonumber\\
&-&
cn(\bfr)^{1/3} \frac{\kappa\gamma}{g(\bfr)^2} \: h(\bfr,\bfr')
\nonumber\\
&-&
c n(\bfr)^{-7/3}\delta(\bfr - \bfr')\:
\nabla \left[\frac{2\kappa\gamma}{g(\bfr)^2} \: \right]\cdot \nabla n(\bfr)
\nonumber\\
&-&
\frac{3c}{4} n(\bfr)^{-4/3}\:
\nabla n(\bfr) \cdot
\nabla\left(
\frac{4\kappa\gamma^2}{g(\bfr)^3}\:h(\bfr,\bfr')\right)
\nonumber\\
&+&
\frac{3c}{4} n(\bfr)^{-4/3}\:
\nabla \left[\frac{2\kappa\gamma}{g(\bfr)^2} \right] \cdot\nabla \delta(\bfr - \bfr')
\nonumber\\
&-&
cn(\bfr)^{-7/3}\delta(\bfr - \bfr')\:
\frac{2\kappa\gamma}{g(\bfr)^2}\:\nabla^2 n(\bfr)
\nonumber\\
&+&
\frac{3c}{4}n(\bfr)^{-4/3}\:
\frac{2\kappa\gamma}{g(\bfr)^2}\:
\nabla ^2 \delta(\bfr - \bfr')
\nonumber\\
&-&
\frac{3c}{4}n(\bfr)^{-4/3}\:\nabla^2 n(\bfr)
\frac{4\kappa\gamma^2}{g(\bfr)^3} \: h(\bfr,\bfr') \: , \label{A14}
\end{eqnarray}
where we defined
\begin{equation}
h(\bfr,\bfr') = \frac{2\nabla n(\bfr) \cdot \nabla \delta(\bfr - \bfr')}{n(\bfr)^{8/3}}
- \frac{8|\nabla n(\bfr)|^2}{3n(\bfr)^{11/3}} \: \delta(\bfr - \bfr').
\end{equation}

To calculate excitation energies, one needs matrix elements of the exchange kernel.
We here consider the case of quantum wells where everything becomes a function of $z$ and $z'$, and
we limit ourselves to intersubband excitations in the quasi-2D limit. Then, only the following matrix element is needed:
\begin{equation}
K_{12} = \int dz \int z' \varphi_1(z) \varphi_2(z)f_{\rm x}^{\rm PBE}(z,z')\varphi_1(z') \varphi_2(z').
\end{equation}
With the explicit form (\ref{A14}) of the PBE exchange kernel, and abbreviating $\xi(z) = \varphi_1(z) \varphi_2(z)$,
one obtains
\begin{eqnarray}
K_{12}&=& -\frac{c}{3}\int dz \: \xi(z)^2n(z)^{-2/3}(1 + \kappa) \nonumber\\
&+& \frac{c\kappa}{3}\int dz \: \xi(z)^2\frac{n(z)^{-2/3}}{g(z)} \nonumber\\
&+& 2c\kappa\gamma \int dz \: \xi(z) \frac{\partial}{\partial z}\left(\frac{\xi(z) n'(z)}{g(z)^2 n(z)^{7/3}}\right) \nonumber\\
&+&\frac{8c}{3}\kappa\gamma\int dz \: \xi(z)^2
\frac{n'(z)^2 }{n(z)^{10/3}g(z)^2} \nonumber\\
&-&2c\kappa\gamma \int dz \: \xi(z)^2 n(z)^{-7/3}n'(z)
\frac{\partial}{\partial z}\left(\frac{1}{g(z)^2}\right) \nonumber\\
&-&6c\kappa\gamma^2\int dz \: \xi(z) \nonumber\\
&&{}\times
\frac{\partial}{\partial z}\left(\frac{n'(z)\frac{\partial}{\partial z}( \xi(z)n'(z)n(z)^{-4/3})}{g(z)^3n(z)^{8/3}} \right)\nonumber\\
&-&8c\kappa\gamma^2\int dz \: \xi(z)\frac{n'(z)^2}{n(z)^{11/3}g(z)^3}\nonumber\\
&&{}\times
\frac{\partial}{\partial z}( \xi(z)n'(z)n(z)^{-4/3}) \nonumber\\
&-&\frac{3c}{2}\kappa\gamma\int dz \: \xi(z)\frac{\partial}{\partial z}
\left[ \xi(z)n(z)^{-4/3}\frac{\partial}{\partial z} \left(\frac{1}{g(z)^2} \right)\right] \nonumber\\
&-&2c\kappa\gamma\int dz \: \xi(z)^2
\frac{n(z)^{-7/3}}{g(z)^2}\ n''(z)\nonumber\\
&+& \frac{3c}{2}\kappa\gamma\int dz \: \xi(z)
\frac{\partial^2}{\partial z^2}\left( \frac{ \xi(z) n(z)^{-4/3} }{g(z)^2}\right) \nonumber\\
&+& 6c\kappa\gamma^2\int dz \: \xi(z)
\frac{\partial}{\partial z}\left( \frac{\xi(z)n''(z)n'(z)}{n(z)^4 g(z)^3}\ \right) \nonumber\\
&+& 8c\kappa\gamma^2\int dz \: \xi(z)^2\frac{n''(z)(n'(z))^2 }{n(z)^5 g(z)^3} \:.
\end{eqnarray}

\section{The PGG kernel for quasi-2DEGs}

In a quantum well of finite width, the single-particle orbitals have the form
\begin{equation}\label{A4}
\varphi_j(\bfr)=e^{i\bfqp \cdot \bfrp} \varphi_j(z) \:,
\end{equation}
where we ignore the normalization factor $A^{-1/2}$ for simplicity.
The PGG exchange kernel (\ref{PGGkernelr}) becomes
\begin{eqnarray}
\lefteqn{
f_{\rm x}\PGG(\bfr,\bfr') = -\frac{2}{|\bfr - \bfr'| n(z) n(z')}}
\nonumber\\
&\times&
\left| \sum_{j=1}^{N_{\rm occ}} \varphi_j(z) \varphi_j(z')\sum_{\bfkp} \theta(k_j-\kp) e^{i\bfkp\cdot(\bfrp -  \bfrp')}
\right|^2
\end{eqnarray}
where $k_j = \sqrt{2(\varepsilon_F - \varepsilon_j)}$. Carrying out the integral over $\bfkp$, and defining
$\bfrhop = \bfrp - \bfrp'$, one finds
\begin{eqnarray}
f_{\rm x}\PGG(\bfr,\bfr') &=& -\frac{2}{|\bfr - \bfr'| n(z) n(z')}
\nonumber\\
&\times&
\left| \sum_{j=1}^{N_{\rm occ}} \varphi_j(z) \varphi_j(z') \frac{k_j J_1(k_j \rhop)}{2\pi \rhop}
\right|^2,
\end{eqnarray}
where $J_1$ denotes a standard Bessel function. Fourier transformation with respect to $\bfrhop$ yields
\begin{eqnarray} \label{PGGkernel}
\lefteqn{\hspace{-1cm}
f_{\rm x}\PGG(\qp,z,z') = - \sum_{j,l}^{N_{\rm occ}}k_j k_l \, \frac{\varphi_j(z)\varphi_l(z) \varphi_j(z')\varphi_l(z')}
{\pi n(z) n(z')}}
\nonumber\\
&\times&
\int_0^\infty d\rhop \: \frac{J_0(\qp \rhop) J_1(k_j \rhop) J_1(k_l \rhop)}{\rhop \sqrt{\rhop^2 + (z-z')^2}}\:.
\end{eqnarray}
If only the first subband is occupied, this simplifies to
\begin{equation} \label{PGGkernel1}
f_{\rm x}\PGG(\qp,z,z') = - \frac{2}{N_s}
\int_0^\infty d\rhop \: \frac{J_0(\qp \rhop)J_1^2(k_1 \rhop)}{\rhop \sqrt{\rhop^2 + (z-z')^2}}\:.
\end{equation}
In the limit of a pure 2DEG, the PGG exchange kernel thus becomes
\begin{equation} \label{PGGkernel2D}
f_{\rm x,\rm 2D}\PGG(\qp) = -\frac{2}{n_{\rm 2D}}\int_0^\infty \frac{d\rhop}{\rhop^2} \: J_0(\qp \rhop) J_1^2(k_F^{\rm 2D}\rhop) \:.
\end{equation}
Let us mention that the PGG exchange kernel (\ref{PGGkernelr}) can also be written as
\begin{equation}
f_{\rm x}^{\rm PGG}(\bfr,\bfr') = 2 \: \frac{g_0(\bfr,\bfr') - 1}{|\bfr - \bfr'|} \:,
\end{equation}
where $g_0(\bfr,\bfr')$ is the noninteracting pair correlation function.
One then finds the following alternative form of the PGG exchange kernel for a 2DEG:
\begin{equation}
f_{\rm x,\rm 2D}\PGG(\qp) = -\frac{\pi}{\qp}\: G_{\ua\ua}^{\rm S}(\qp) \:,
\end{equation}
where
\begin{eqnarray}
G_{\ua\ua}^{\rm S}(\qp) &=& -\frac{\qp}{2\pi^2 n}\int \frac{d^2\qp'}{|\bfqp - \bfqp'|}\:  [S_0(\qp')-1]
\\
&=&
-\frac{2\qp}{\pi^2 n} \int_0^\infty\!  \frac{\qp' \,d\qp'}{\qp + \qp'}K\! \left(\frac{\sqrt{4\qp \qp'}}{\qp + \qp'}\right)\!
[S_0(\qp')-1]
\nonumber\\
&&
\end{eqnarray}
is the so-called Slater local field factor ($S_0$ is the noninteracting static structure factor
and $K$ is the complete elliptic integral of the first kind).\cite{GV}

\section{ISTLS in the exchange-only limit}

\begin{figure}[t]
\begin{center}
\includegraphics[angle=0, width=8.5cm]{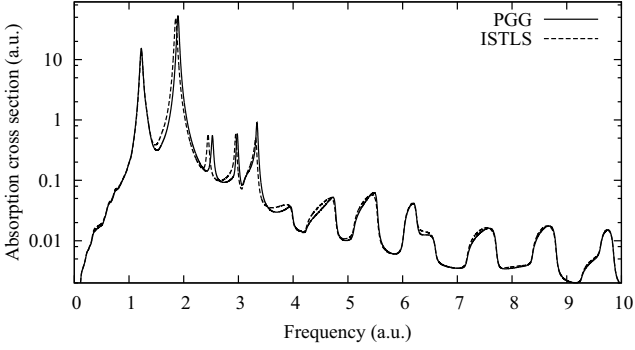}
\end{center}
\caption{Photoabsorption cross section for $\qp=0$ intersubband charge plasmons, for a quantum well with 5 occupied subbands,
comparing PGG and exchange-only ISTLS.}
\label{fig9}
\end{figure}

In the inhomogeneous STLS (ISTLS) approach, the xc kernel has the following tensorial form: \cite{Dobson2002,Dobson2009}
\begin{equation}
f_{\rm xc,\mu\nu}\ISTLS(\bfr,\bfr') = -\frac{2}{\omega^2}[g(\bfr,\bfr')-1] \frac{\partial}{\partial_\mu}\frac{1}{|\bfr - \bfr'|}
\frac{\partial}{\partial_\nu'} \:,
\end{equation}
where $\mu,\nu$ denote Cartesian coordinates and $g(\bfr,\bfr')$ is the pair correlation function.
The exchange-only limit of this expression is obtained by using the noninteracting pair correlation function, which yields
\begin{equation}\label{B3}
f_{\rm x,\mu\nu}\ISTLS(\bfr,\bfr') =  2
\frac{\left|\sum_{j=1}^{N_{\rm occ}} \varphi_{j}(\bfr) \varphi^*_{j}(\bfr')\right|^2}
{ \omega^2n(\bfr) n(\bfr')}\frac{\partial}{\partial_\mu}\frac{1}{|\bfr-\bfr'|}\frac{\partial}{\partial_\nu'}\:.
\end{equation}
We consider the case of a quantum well with finite width, where the Kohn-Sham orbitals have the form (\ref{A4}),
and we limit ourselves to plasmon modes with in-plane wavevector $\qp=0$, so that the dynamics is uniform within the
plane of the well and, hence, effectively one-dimensional. Then, only the $zz$ component of the tensorial xc kernel is relevant, and it is straightforward to transform it to a scalar exchange kernel.\cite{Ullrich2012}
Using the same notation as in Appendix B, we obtain
\begin{eqnarray}
f_{\rm x}^{\rm ISTLS}(\qp=0,z,z')
&=&
\int_z^\infty dz_1 \int_0^\infty \frac{d\rhop}{\rhop} \nonumber\\
&\times&
\frac{\left| \sum_j^{N_{\rm occ}} \varphi_j^*(z_1) \varphi_j(z') \: k_j J_1(k_j\rhop)\right|^2}{\pi  n(z_1) n(z')}
\nonumber\\
&\times&
\left( \frac{\partial}{\partial z_1}\frac{1}{\sqrt{\rhop^2 + (z_1-z')^2}}\right) .
\end{eqnarray}
Comparing with Eq. (\ref{PGGkernel}) [notice that $J_0(0)=1$], we can rewrite this as
\begin{eqnarray}
f_{\rm x}^{\rm ISTLS}(0,z,z')
&=&
f_{\rm x}^{\rm PGG}(0,z,z') \nonumber\\
&-&
\int_z^\infty dz_1
\int_0^\infty \frac{d\rhop}{\rhop}\nonumber\\
&\times&
\sum_{l,m}^{N_{\rm occ}}\frac{k_l k_m J_1(k_l\rhop)J_1(k_m\rhop)}{\sqrt{\rhop^2 + (z_1-z')^2}}
\nonumber\\
&\times&
\frac{\partial}{\partial z_1\! }\left(\!
\frac{ \varphi_l(z_1) \varphi_l^*(z')\varphi_m^*(z_1) \varphi_m(z')}{\pi n(z_1) n(z')}\!\right) \quad
\end{eqnarray}
It thus turns out that the ISTLS exchange kernel is equal to the PGG exchange kernel plus a correction term.
If only the lowest subband is occupied ($N_{\rm occ}=1$), the correction term vanishes because then the derivative with
respect to $z_1$ gives zero.

Figure \ref{fig9} gives a comparison of PGG and ISTLS for the case of a quantum well with 5 occupied subbands.
The figure shows the frequency-dependent photoabsorption cross section corresponding the intersubband charge plasmons.
As can be seen, the difference between PGG and ISTLS is marginal.


\begin{thebibliography}{99}
\bibitem{footnote}
The usual Coulomb interaction, $1/|\bfr - \bfr'|$, applies in 3D and 2D, but in 1D it would lead to
divergencies and requires regularization (see Ref. \onlinecite{GV}).
\bibitem{2Dfootnote}
In this paper we distinguish quasi-2D and strictly 2D systems. By quasi-2D we mean the quantum limit
in which all electrons occupy the lowest subband in the well, but the electron density distribution still has
a finite lateral width; strictly 2D implies the limit in which the width has become infinitesimal.
\end{thebibliography}
\end{document}